\documentclass[12pt,preprint]{aastex}

\slugcomment{}
\shorttitle{}
\shortauthors{}

\begin{document}

\title{Simultaneous Determination of the Cosmic Ray Ionization Rate and
Fractional Ionization in DR21(OH)}

\author{Talayeh Hezareh\altaffilmark{1},} 
\email{hezareh@astro.uwo.ca}
\author{Martin
Houde\altaffilmark{1}, Carolyn M$^{\mathrm{c}}$Coey \altaffilmark{1,2},
Charlotte Vastel\altaffilmark{3}, \and Ruisheng Peng\altaffilmark{4}}

\altaffiltext{1}{The Department of Physics and Astronomy, The University of Western Ontario, London, Ontario, Canada N6A 3K7}

\altaffiltext{2}{Department of Physics and Astronomy, University of Waterloo, 200 University Avenue W., Ontario, Canada N2L 3G1}
\altaffiltext{3}{Centre d'Etude Spatiale des Rayonnements, 9 avenue du colonel Roche BP 44346, 31028 Toulouse Cedex 4, FRANCE}
\altaffiltext{4}{Caltech Submillimeter Observatory, 111 Nowelo Street, Hilo, HI 96720}

\begin{abstract}
We present a new method for the simultaneous calculation of the cosmic
ray ionization rate, $\zeta_{\mathrm{H}_{2}}$, and the ionization
fraction, $\chi_{\mathrm{e}}$, in dense molecular clouds. A simple
network of chemical reactions dominant in the creation and destruction
of HCNH$^+$ and HCO$^+$ is used in conjunction with observed pairs
of rotational transitions of several molecular species in order to
determine the electron abundance and the H$_{3}$$^{+}$ abundance.
The cosmic ray ionization rate is then calculated by taking advantage
of the fact that, in dark clouds, it governs the rate of creation
of H$_{3}$$^{+}$. We apply this technique to the case of the star-forming
region DR21(OH), where we successfully detected the ($J=3 \rightarrow 2$)
and ($J=4 \rightarrow 3$) rotational transitions of HCNH$^+$. We
also determine the C and O isotopic ratios in this source to be $^{12}\mathrm{C/^{13}\mathrm{C}}=63\pm4$
and $^{16}\mathrm{O/^{18}\mathrm{O=318\pm64}}$, which are in good
agreement with previous measurements in other clouds. The significance
of our method lies in the ability to determine $N(\mathrm{H_{3}^{+}})$
and $\chi_{\mathrm{e}}$ directly from observations, and estimate
$\zeta_{\mathrm{H}_{2}}$ accordingly. Our results, $\zeta_{\mathrm{H}_{2}}=3.1\times10^{-18}\:\mathrm{s}^{-1}$
and $\chi_{\mathrm{e}}=3.2\times10^{-8}$, are consistent with recent
determinations in other objects. 
\end{abstract}


\section{Introduction}

The ionization fraction (or the electron abundance), $\chi_{\mathrm{e}}\equiv\mathrm{\mathit{n}(e)/\mathit{n}(H_{2})}$,
where $n(\mathrm{e})$ is the electron number density and $\mathit{n}(\mathrm{H_{2}})$
the hydrogen molecule number density, plays a key role in the chemistry
and dynamics of dense molecular clouds. Due to the low temperatures
within dense clouds, the chemistry is dominated by reactions between
ions and neutral species. Furthermore, if a magnetic field permeates
the cloud, the ions interact directly with the field through the Lorentz
force, while the rate of collisions between the ions and the neutral
species determines the degree with which the neutral gas couples to
the magnetic field. The ionization fraction therefore determines the
strength of this coupling, and hence the ambipolar diffusion timescale
in a molecular cloud, which is a measure of the stability of the cloud
against gravitational collapse \citep{shu}. Moreover, determination
of the fractional ionization can be essential in precisely estimating
the magnetic strength in molecular clouds \citep{li}.

In order to determine the ionization fraction in a cloud, it is important
to identify the main ionization processes in the region under study.
Star formation occurs in dense cores, which are regions of high extinction
where self-shielding prevents the UV photo-ionization of H$_{2}$.
It is also expected that X-ray ionization is only significant in the
vicinity of strong X-ray sources such as active galactic nuclei \citep{mccall}
and OB stars. Therefore, cosmic ray ionization is believed to dominate
photo-ionization in dense cores \citep{mckee}. 

Cosmic rays also heat the interstellar gas and drive interstellar
chemistry in dense molecular clouds. Direct determination of the cosmic
ray ionization rate, $\zeta_{\mathrm{H}_{2}}$, is achievable by studying
the abundance of H$_{3}$$^{+}$ due to the relative simplicity of
its chemistry \citep{mccall}. This important molecule is created
via cosmic ray ionization of the H$_{2}$ molecule, and is highly
reactive with electrons and neutral species present in such clouds. For example,
it reacts with HCN, CO and N$_{2}$ to form HCNH$^{+}$, HCO$^{+}$
and N$_{2}$H$^{+}$, respectively. The resulting ions react with different
neutral molecules, producing other ionic or neutral species. For this
study, we assume that H$_{3}$$^{+}$ is mainly produced by cosmic
rays. 

Various techniques have been developed in the past to estimate the
ionization fraction and cosmic ray ionization rate. For the prior,
there have been studies that based its determination on measurements
of the degree of deuterium fractionation through DCO$^{+}$ and HCO$^{+}$
abundance ratios (e.g., \citealt{caselli}). The application of this
technique in cold clouds is limited, since the freeze-out of molecules
onto the grain surfaces affects the degree of deuterium fractionation
independently of the ionization fraction \citep{caselli}. Observations
of H$_{3}$$^{+}$ absorption lines \citep{mccall} together with
appropriate chemical models have been used for the determination of
$\zeta_{\mathrm{H}_{2}}$ in diffuse clouds and envelopes of molecular
clouds \citep{vanDer,indriolo}. However, this method is not applicable
to cold starless cores since H$_{3}$$^{+}$ has no electric dipole
moment; it cannot be used to trace such regions. 

In dense cores, typical values for $\zeta_{\mathrm{H}_{2}}$ are estimated
to be $(1-5)\times10^{-17}$ s$^{-1}$ \citep{dalgarno}, while the
fractional ionization is found to lie within a small range, $10^{-7.3}\lesssim\chi_{\mathrm{e}}\lesssim10^{-6.9}$
\citep{bergin}. However, some recent studies have found lower values
for these parameters in dark clouds. For example, \citet{maret} obtained
a fractional ionization of $5\times10^{-9}$ with respect to H nuclei
corresponding to a cosmic ray ionization of $(1-6)\times10^{-18}$
s$^{-1}$ in the Barnard 68 prestellar core, while \citet{flower}
reported $n_{\mathrm{e}}/n_{\mathrm{H}}=1\times10^{-8}$ and $\zeta_{\mathrm{H}_{2}}=2\times10^{-18}$
s$^{-1}$ in TMC-1. Since the two parameters are interdependent, one
is usually determined by restricting the other through fitting theoretical
models to observations (see for example, \citealt{wootten}; \citealt{plume}). 

In this work we show that it is possible to estimate the cosmic ray
ionization rate and fractional ionization simultaneously, using observational
spectroscopic data together with a simple network of chemical reactions
that involve H$_{3}$$^{+}$ and the electron abundance. These reactions
are responsible for the formation and destruction of well-studied
molecular species that co-exist in dense molecular clouds. Accordingly,
we chose to study HCNH$^{+}$ and HCO$^{+}$ and applied this technique
to the star-forming region DR21(OH).

We explain the observational procedure in $\S 2$ and present our
technique in $\S 3$. Our numerical calculations and results are described
in $\S 4$, followed by a discussion and summary in $\S 5$.

\section{Observations}

\subsection{Source description}

DR21(OH) ($\alpha=20^{\mathrm{h}}39^{\mathrm{m}}01^{\mathrm{s}}$
and $\delta=42^{\circ}22^{'}37.7^{''}$, J2000) is located about $3\arcmin$
to the north of DR21, a well-known massive star formation site in
the Cygnus X region that lies at a distance of approximately $3$
kpc \citep{genzel}. Also known as W75S, it is made up of several
compact sources, namely DR21(OH)Main, DR21(OH)N, DR21(OH)S and DR21(OH)W,
which are all active star-forming regions \citep{curran}. This work
is focused on the brightest component, DR21(OH)Main. Continuum studies
have detected two dense cores, MM1 and MM2, with a total mass of about
$125\mathrm{\: M}_{\odot}$ in the center of DR21(OH)Main \citep{Woody}.
This source has been extensively studied in the infrared and also
mapped in CO (\citealt{dickel}; \citealt{magnum}), and no young
stars with strong radiation fields have been observed around it \citep{davis}.
Also, no centimeter-wavelength continuum sources have been detected
within the source, suggesting a lack of HII regions \citep{johnston},
implying that it is in an early stage of evolution and thus suitable
for our analysis.

\subsection{Spectroscopic data}

We obtained all observations at the Caltech Submillimeter Observatory
(CSO), located on Mauna Kea, Hawaii. In Table \ref{ta:data}, we list
the detections of different rotational transitions from the molecular
species that form the main creation and destruction routes for HCNH$^+$
and HCO$^+$ (see $\S 3$). Among these molecules, HCO$^{+}$, CO,
HCN and HNC were optically thick, and therefore not suitable for our
analysis. Instead, we observed the optically thin isotopologues H$^{13}$CO$^{+}$,
H$^{13}$CN and HN$^{13}$C, as well as $^{12}$C$^{18}$O, $^{13}$C$^{18}$O,
and $^{13}$C$^{16}$O in order to determine the $^{12}$C/$^{13}$C
and $^{16}$O/$^{18}$O isotopic ratios needed for the calculation
of the abundance of the main species from their observed isotopologues. 

In October 2006, we detected HCNH$^+$ in the $J=3\rightarrow2$ and
$J=4\rightarrow3$ transitions in DR21(OH)Main using the 200-300 GHz
receiver. For these observations, standard telescope efficiencies
of 66 \% at 222 GHz (beam width $\sim $33$''$) and 60 \% at 296 GHz
(beam width $\sim$ 25$''$) were used. Data were taken in a position
switching mode and the pointing was checked regularly using scans
on Uranus. H$^{13}$CO$^{+}$ and H$^{13}$CN were observed during
the months of October, November and December 1999, using the 200-300
GHz and 300-400 GHz receivers. These spectra were calibrated using
scans made on planets available during that period (Mars, Jupiter
and Saturn). The telescope efficiencies were $\sim70$ \% for the
200-300 GHz receiver (beam width of $\sim32\arcsec$) and $\sim60$
\% for the 300-400 GHz receiver (beam width of $\sim20\arcsec$) \citep{houde}. 

\begin{figure}
\epsscale{0.7}\plotone{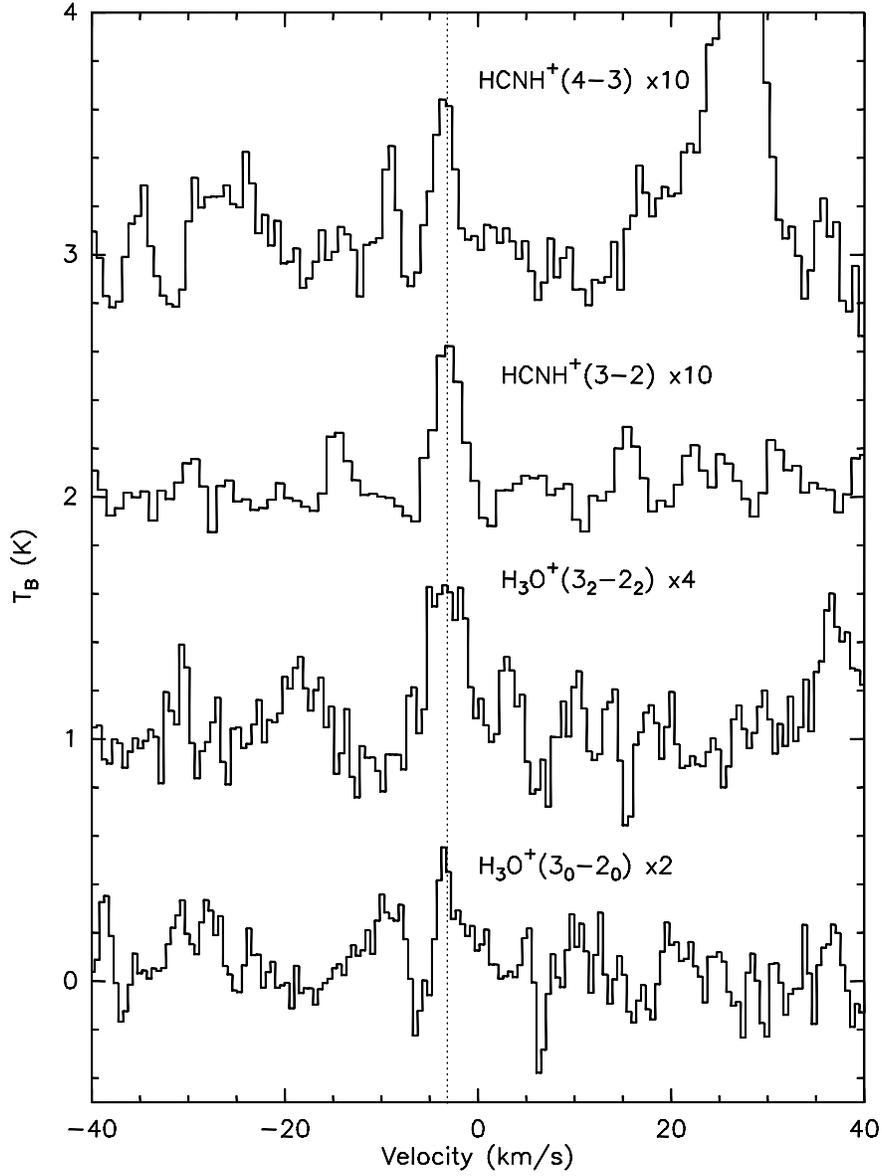}\caption{\label{fig1} The spectra for the HCNH$^{+}$ transitions $J=4\rightarrow3$
and $J=3\rightarrow2$, and para H$_{3}$O$^{+}$ ($J=3_{2}\longrightarrow2_{2}$)
and ortho H$_{3}$O$^{+}$ ($J=3_{0}\longrightarrow2_{0}$) towards
the center position of DR21(OH). }

\end{figure}

\begin{figure}
\epsscale{0.7}\plotone{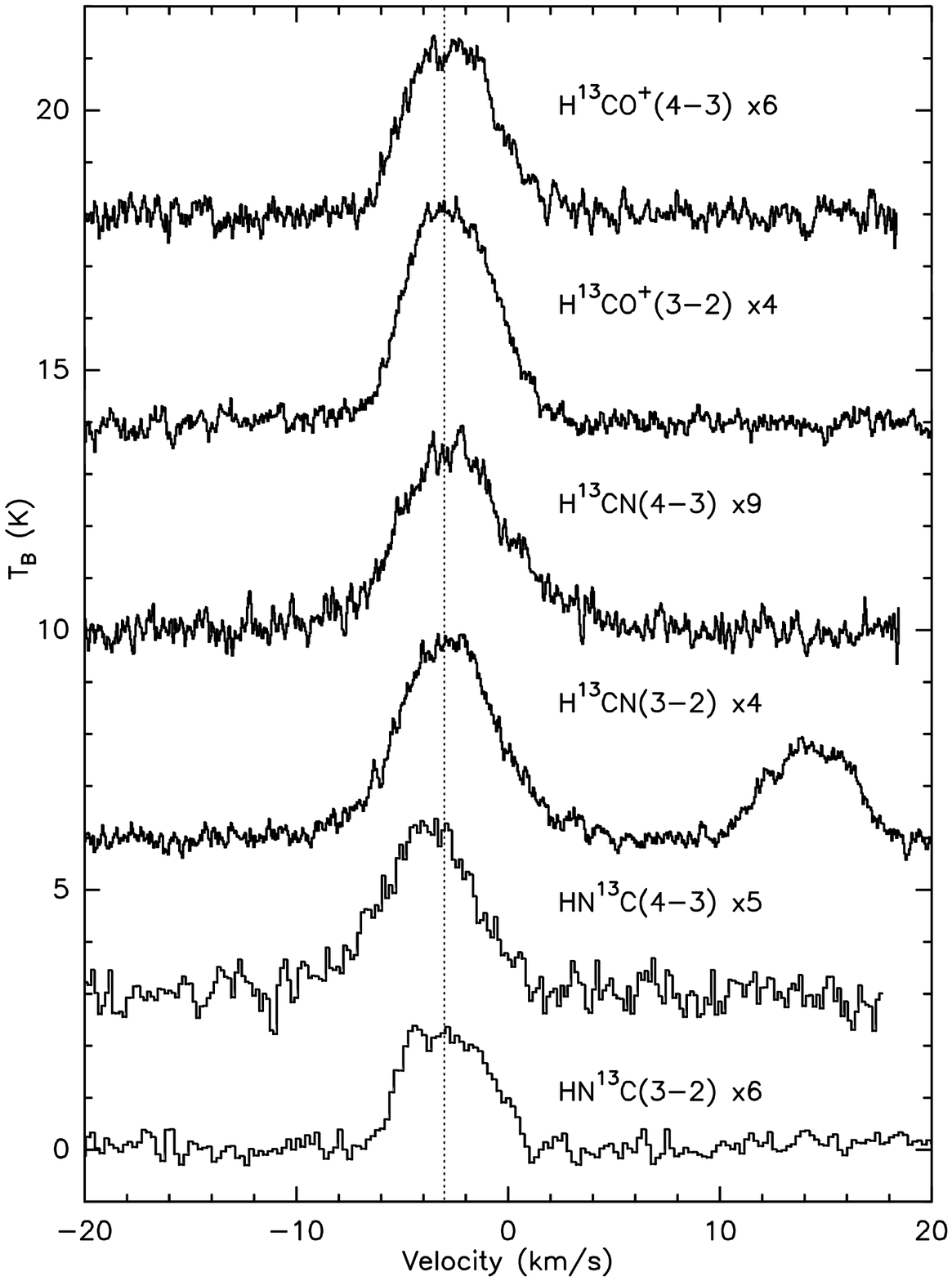}

\caption{\label{fig2} The spectra for the $J=4\rightarrow3$ and $J=3\rightarrow2$
transitions of H$^{13}$CN, HN$^{13}$C, and H$^{13}$CO$^{+}$ towards
the center position of DR21(OH). }

\end{figure}

\begin{figure}
\epsscale{0.7}\plotone{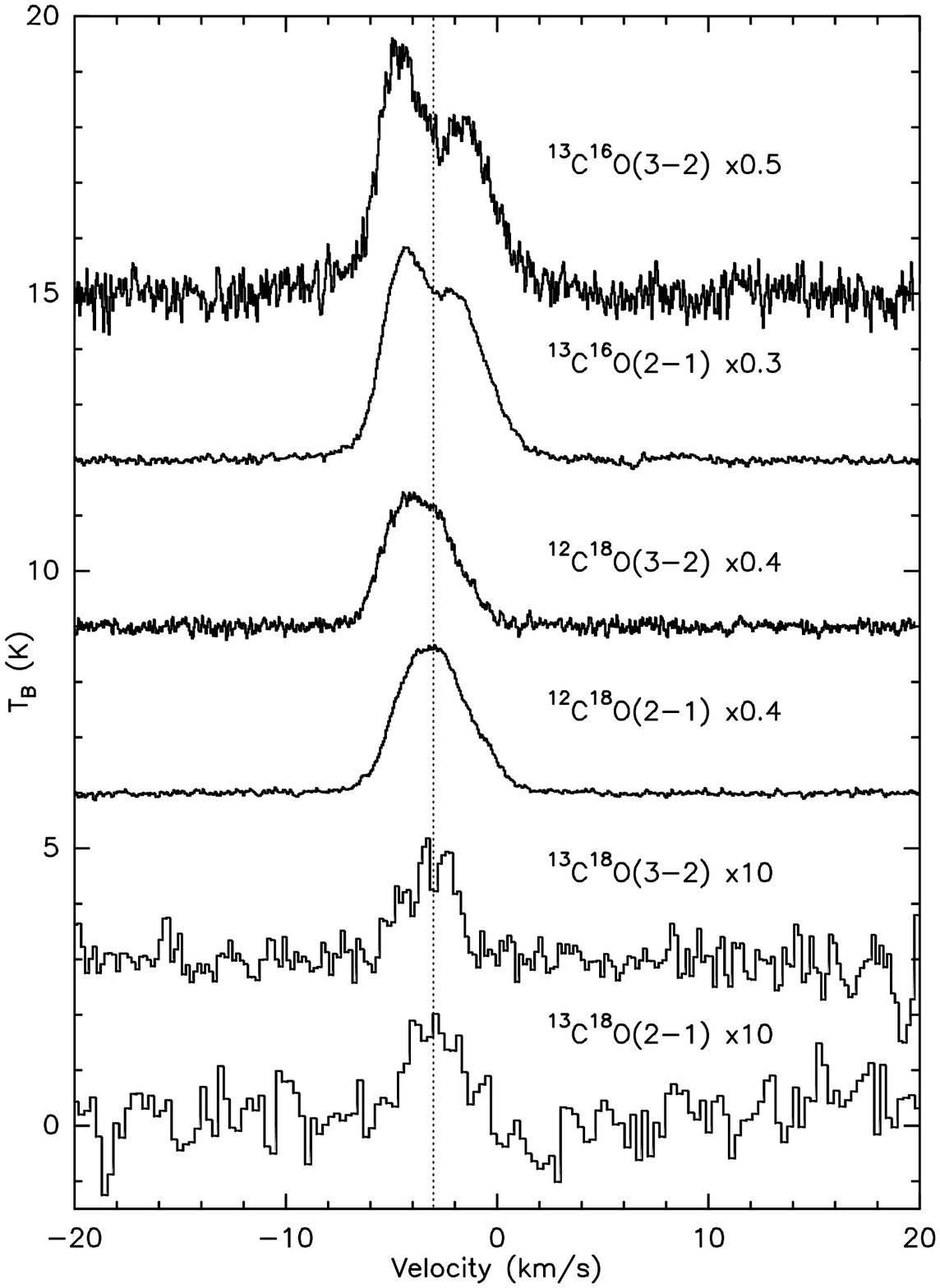}

\caption{\label{fig3} The spectra for the $J=3\rightarrow2$ and $J=2\rightarrow1$
transitions of $^{13}$C$^{18}$O, $^{12}$C$^{18}$O, and $^{13}$C$^{16}$O
towards the center position of DR21(OH). }

\end{figure}

We obtained the final set of observational data required for this
analysis in October and November 2007. During that period, the $^{12}$C$^{18}$O,
$^{13}$C$^{18}$O, and $^{13}$C$^{16}$O molecular species were
detected in two rotational transitions ($J=2\rightarrow1$ and $J=3\rightarrow2$)
towards the center position of DR21(OH). We also attempted to detect
H$_{3}$O$^{+}$ in $J=3_{0}\rightarrow2_{0}$ and $J=3_{2}\rightarrow2_{2}$
transitions at 396 GHz and 364 GHz, respectively. Although we were
successful in detecting the 364 GHz transition, we did not record
an acceptable detection for the 396 GHz transition. Standard telescope
efficiencies were also used for these sets of data (i.e., $\sim66$
\% at 200-300 GHz with a beam width of $\sim33\arcsec$, and $\sim58$
\% at 300-400 GHz with a beam width of $\sim22\arcsec$). Figure \ref{fig1}
shows the spectra for HCNH$^{+}$ and H$_{3}$O$^{+}$. The spectra
for the $J=3\rightarrow2$ and $J=4\rightarrow3$ transitions of H$^{13}$CO$^{+}$,
H$^{13}$CN and HN$^{13}$C are shown in Figure \ref{fig2}, while
the detections of the CO isotopologues are presented in Figure \ref{fig3}.
All the data reduction was carried out using the GILDAS%
\footnote{URL: http://iram.fr/IRAMFR/GILDAS/%
} and CASSIS%
\footnote{URL: http://cassis.cesr.fr/%
} software packages.

DR21(OH) is well known to have a multiply peaked structure, suggesting
that the cloud is in an early stage of massive star formation \citep{Richardson1}.
Therefore, it is important to determine whether or not the emission
from the molecules under study arises from the same volume of gas
along the line of sight. In order to confirm the coexistence of the
molecular species along each line of sight, it is useful to obtain
velocity maps and analyze them together with line profiles. However,
velocity mapping, especially for the faint lines of HCNH$^{+}$ and
H$_{3}$O$^{+}$, required a longer integration time than was available
to us. Nevertheless, consideration of the spectra in Figures \ref{fig1},
\ref{fig2} and \ref{fig3} shows an alignment of the peak velocities
of the detected lines near -3 km s$^{-1}$. Therefore, to a good approximation,
the observed molecular species are coexistent in the region along
our line of sight. Notable exceptions are the $J=4\rightarrow3$ transition
of HN$^{13}$C and $J=3\rightarrow2$ transition of $^{12}$C$^{18}$O
in Figure \ref{fig2} and \ref{fig3}, respectively. The corresponding
observations are, therefore, not used in our analysis.

\placetable{ta:data}

\section{Method}

In this section, we describe the main creation and destruction paths
of HCNH$^{+}$ and HCO$^{+}$ in dense clouds and the method used
to calculate $n(\mathrm{e})$, $n(\mathrm{H_{3}^{+}})$ and $\zeta_{\mathrm{H_{2}}}$.
In dense clouds, the chemistry is dominated by ion-neutral reactions
and the main formation reactions for HCNH$^{+}$ are as follows \citep{schilke}

\begin{eqnarray}
\mathrm{H}_{3}^{+}+\mathrm{HCN}\left(\mathrm{HNC}\right) & \longrightarrow & \mathrm{HCNH}^{+}+\mathrm{H}_{2}\label{eq:my1}\\
\mathrm{HCO}^{+}+\mathrm{HCN}\left(\mathrm{HNC}\right) & \longrightarrow & \mathrm{HCNH}^{+}+\mathrm{CO}\label{eq:my2}\\
\mathrm{H}_{3}\mathrm{O}^{+}+\mathrm{HCN}\left(\mathrm{HNC}\right) & \longrightarrow & \mathrm{HCNH}^{+}+\mathrm{H}_{2}\mathrm{O}.\label{eq:my3}
\end{eqnarray}

\noindent Likewise, in dark clouds, HCO$^{+}$ is mainly formed by
the reaction of H$_{3}$$^{+}$ with CO (\citealt{herbst,watson}) 

\begin{equation}
\mathrm{H_{3}^{+}+CO\longrightarrow HCO^{+}+H_{2}}.\label{eq:my4}\end{equation}

\noindent Both HCNH$^{+}$ and HCO$^{+}$ are mainly removed through
dissociative recombination with electrons (\citealt{schilke}; \citealt{plume})

\begin{eqnarray}
\mathrm{HCNH^{+}+e^{-}} & \longrightarrow & \mathrm{HCN+H}\label{eq:hcnh+}\\
 & \longrightarrow & \mathrm{HNC+H}\nonumber \\
 & \longrightarrow & \mathrm{CN+2H}\nonumber \\
\mathrm{HCO^{+}+e^{-}} & \longrightarrow &  \mathrm{CO+H}\label{eq:hco2}
\end{eqnarray}

\noindent The rate coefficients of the above reactions are taken
from the UMIST database \citep{Woodall} and are listed in Table \ref{ta:ki}.
The kinetic temperature in dark molecular clouds is usually between
10 to 40 K; we follow \citet{wilson} in adopting a kinetic temperature
of $20$ K for DR21(OH) to evaluate these rate coefficients. 

\placetable{ta:ki}

There are several other reactions that can contribute to the creation
and destruction of HCNH$^{+}$ and HCO$^{+}$ in dense clouds, but
they can safely be neglected due to their low rate coefficients and/or
the relatively low abundance of the molecules involved (e.g., reactions
involving C$_{2}$H$_{2}$$^{+}$, H$_{2}$O$^{+}$, H$_{2}$CO$^{+}$,
HNO$^{+}$, C$_{2}$$^{+}$, and H$_{2}$S$^{+}$). 

We can use the aforementioned reactions to equate the rates of formation
and destruction of HCNH$^{+}$ (equations (\ref{eq:my1})-(\ref{eq:my3}),
and (\ref{eq:hcnh+})) and HCO$^{+}$ (equations (\ref{eq:my4}) and
(\ref{eq:hco2})) to obtain the following expressions for the electron
and H$_{3}$$^{+}$ abundances 

\begin{equation}
\mathrm{\mathit{n}(\mathrm{\mathit{\mathrm{e}}})=\frac{[\mathit{n}(\mathrm{HCN})+\mathit{n}(\mathrm{HNC})][\mathit{\mathrm{\mathit{n}(H_{3}^{+})\mathit{k_{1}}+}n}(HCO^{+})\mathit{k_{2}+\mathrm{\mathit{n}(H_{3}O^{+})\mathit{k_{3}}}}]}{\mathit{n}(HCNH^{+})\mathit{k_{5}}}},\label{eq:n(e)1}\end{equation}

\begin{equation}
n(\mathrm{H}_{3}^{+})=\frac{n(\mathrm{HCO^{+}})n(\mathrm{\mathit{\mathrm{e}}})k_{6}}{n(\mathrm{CO})k_{4}}\:.\label{eq:h3+}\end{equation}

\noindent Since we have obtained all the observations necessary to
determine the abundance of the HCN, HNC, HCO$^{+}$, H$_{3}$O$^{+}$,
HCNH$^{+}$, and CO molecular species, we are left with a set of two
equations and two unknowns (i.e., $n$(H$_{3}$$^{+}$) and $n$(e)).
It will therefore be straightforward to simultaneously determine $n$(H$_{3}$$^{+}$)
and $n$(e) using equations (\ref{eq:n(e)1}) and (\ref{eq:h3+})
and our spectroscopic data. 

The cosmic ray ionization rate can be obtained directly from the abundance
of H$_{3}$$^{+}$, which forms as follows (\citealt{solomon}; \citealt{bowers})

\begin{eqnarray}
\mathrm{H_{2}+CR} & \longrightarrow & \mathrm{H}_{2}^{+}+\mathrm{e}^{-}\label{eq:h3+1}\\
\mathrm{H_{2}^{+}}+\mathrm{H}_{2} & \longrightarrow & \mathrm{H_{3}^{+}+H}\:.\label{eq:h3+2}
\end{eqnarray}

The reaction between $\mathrm{H_{2}^{+}}$ and $\mathrm{H}_{2}$ proceeds
very rapidly \citep{solomon} and is limited by the abundance of $\mathrm{H_{2}^{+}}$.
The rate of formation of H$_{3}$$^{+}$ is therefore governed by
the ionization of $\mathrm{H}_{2}$ by cosmic rays. 

The main destruction path for H$_{3}$$^{+}$ in dense cores is through
reactions with CO \citep{mccall}, due to the high reaction rate and
the fact that the latter has the highest fractional abundance with
respect to $\mathrm{H}_{2}$ in molecular clouds. Furthermore, H$_{3}$$^{+}$
is also destroyed by dissociative recombination with electrons through

\begin{eqnarray}
\mathrm{H_{3}^{+}+e^{-}} & \longrightarrow & \mathrm{H_{2}+H}\label{eq:h3}\\
 & \longrightarrow & \mathrm{H+H+H.}\nonumber 
\end{eqnarray}

Although the overall rate for this reaction is an order of magnitude
lower than that involving CO (see equation (\ref{eq:my4})), we still
include it in what follows for completeness. Again, assuming statistical
equilibrium between the formation and destruction rate of H$_{3}$$^{+}$,
we obtain

\begin{equation}
\zeta_{\mathrm{H}_{2}}=\frac{n(\mathrm{H}_{3}^{+})n(\mathrm{CO})k_{4}+n(\mathrm{H}_{3}^{+})n(\mathrm{e})k_{11}}{n(\mathrm{H}_{2})}\:.\label{eq:zeta}\end{equation}

The main advantage of our method is that it enables us to calculate
$\chi_{\mathrm{e}}$ and $\zeta_{\mathrm{H}_{2}}$ simultaneously
through equations (\ref{eq:n(e)1}), (\ref{eq:h3+}) and (\ref{eq:zeta}),
without the need to consider more extensive chemical networks for
each molecular species involved. As mentioned before, this is because
the required abundances are determined directly from observations.
This feature differentiates ours from previous techniques.

\section{Results}

The expressions for $\chi_{\mathrm{e}}$ and $\zeta_{\mathrm{H}_{2}}$
in the previous section are couched in terms of abundance but could
equivalently be written in terms of column density, which is a more
natural quantity to use when analyzing observation. We calculated
the excitation temperature of every observed molecular species (or
at least for one of its isotopologues) that appears in equations (\ref{eq:n(e)1})
and (\ref{eq:h3+}) in order to calculate its column density. The
C and O isotopic ratios in the cloud were determined with observations
of the $^{12}\mathrm{C}^{18}\mathrm{O}$, $^{13}\mathrm{C}^{18}\mathrm{O}$
and $^{13}\mathrm{C}^{16}\mathrm{O}$ molecular species, and were
used to evaluate the column density of the common molecular species
(more below).

\subsection{Excitation temperatures and column densities}

We determined the excitation temperatures and column densities of
the observed molecules using a LTE approach. For this purpose, we
detected two transitions for each molecular species and obtained the
ratio of their integrated intensities (\citealt{blake}; \citealt{emerson}).
The brightness temperature of a source can be written as (\citealt{kutner})

\begin{equation}
T_{b}(\nu)=T_{0}\biggl[\frac{1}{e^{(T_{0}/T_{ex})}-1}-\frac{1}{e^{(T_{0}/T_{\mathrm{CMB}})}-1}\biggr](1-e^{-\tau(\nu)}),\label{eq:Tb}\end{equation}

\noindent where $T_{ex}$ is the excitation temperature, $T_{\mathrm{CMB}}$
the cosmic microwave background brightness temperature (2.7 K), $\tau(\nu)$
the optical depth, and $T_{0}=h\nu/k$, with $\nu$ the frequency.
The term involving the background brightness temperature is insignificant
compared to the other term and is therefore neglected. The mean optical
depth can be written as a function of column density as 

\begin{eqnarray}
\tau\Delta v & \equiv & \int\tau(v)dv\label{eq:tau}\\
 & = & \frac{A_{ul}c^{3}N_{u}}{8\pi\nu^{3}}(e^{(T_{0}/T_{ex})}-1),\nonumber 
\end{eqnarray}

\noindent where $\Delta v$ is the extent of the spectral line, $A_{ul}$
is the Einstein coefficient for spontaneous emission (listed in Table
\ref{ta:data}) and $N_{u}$ is the column density of the upper level
of a transition, which, in turn, can be written as a function of the
integrated brightness temperature and mean optical depth of the line
as 

\begin{equation}
N_{u}=\frac{8\pi k\nu^{2}}{hc^{3}A_{ul}}\Bigl(\frac{\tau}{1-e^{-\tau}}\Bigr)\int T_{b}dv,\label{eq:nu/gu}\end{equation}

\noindent where the term $\int T_{b}dv$ is the integrated intensity
of the spectral line profile and the term $\tau/(1-e^{-\tau})$ (hereafter
abbreviated by $\beta$) is the optical depth correction factor \citep{lp93}.
For optically thin lines, where $\tau\ll1$, $\beta$ approaches unity.
Within the LTE approximation, the population of the levels is assumed
to be thermalized and the total column density of the molecular species
along the observer's line of sight, $N_{tot}$, can be expressed as

\begin{equation}
N_{tot}=\frac{8\pi k\nu^{2}U(T_{ex})}{hc^{3}A_{ul}g_{u}}e^{(E_{u}/kT_{ex})}\beta\int T_{b}dv,\label{eq:Nt}\end{equation}

\noindent where $U(T_{ex})$ is the partition function at the excitation
temperature $T_{ex}$, while $g_{u}$ and $E_{u}$ are the upper state
degeneracy and energy, respectively.

In order to obtain the excitation temperature of a molecular species
that exhibits optically thin line profiles, we obtained the ratio
of the integrated intensities of its two observed transitions using
equation (\ref{eq:Nt}). For such molecules, we adopted $\beta=1$
for the corresponding spectral lines and solved for $T_{ex}$, assuming
it to be the same for both transitions. The optical depths of the
lines were subsequently calculated using equation (\ref{eq:tau}).
For example, in the case of H$^{13}$CN we obtained a $T_{ex}$ of
$13$ K and opacities of $0.09$ and $0.06$ for the $J=3\rightarrow2$
and $J=4\rightarrow3$ transitions, respectively. The optical depths
for all transitions thus calculated are listed in Table \ref{ta:data}.
This method could not be used for $^{12}$C$^{18}$O and $^{13}$C$^{16}$O,
as their transitions are not optically thin. 

Inspection of the $^{12}$C$^{18}$O spectra in Figure \ref{fig3}
reveals that the line profiles are slightly saturated. Furthermore,
the peak velocity of the $J=3\rightarrow2$ transition is clearly
blue-shifted from the $J=2\rightarrow1$ transition and the vast majority
of the other observed lines rendering it unsuitable for this analysis,
since it possibly arises from a different region. Therefore, care
had to be taken toward the determination of the excitation temperature
and optical depth of this molecule. Since the line profile for the
$J=2\rightarrow1$ transition is relatively symmetric and of appropriate
shape, we fitted it with a Gaussian profile. We then used the equation
for a Gaussian line from \citet{vastel} and Newton's method to numerically
solve the equation 

\begin{equation}
N_{u}-\frac{8\pi k\nu^{2}}{hc^{3}A_{ul}}\Bigl(\frac{\tau(N_{u})}{1-e^{-\tau(N_{u})}}\Bigr)T_{b}\delta v=0\label{eq:newton}\end{equation}
\noindent for $N_{u}$ and subsequently for $\tau(N_{u})$ for a
very narrow velocity interval $\delta v$ centered on $v$. The dependency
of the mean optical depth on $N_{u}$ follows from equation (\ref{eq:tau}).
This way, we obtained the optical depth in different parts of the
line profile to model the spectrum using a range of excitation temperatures
($10<T_{ex}<25$). Every optical depth value was converted to a brightness
temperature using equation (\ref{eq:Tb}) and consequently, assuming
a beam filling factor of 1, to an antenna temperature through $T_{A}^{*}=\eta T_{b}$,
where $\eta$ is the telescope efficiency at the appropriate frequency.
The excitation temperature of $^{12}$C$^{18}$O was taken to be the
temperature of the best fit to the $J=2\rightarrow1$ spectrum. The
line profiles of H$_{3}$O$^{+}$ ($J=3_{2}\rightarrow2_{2}$) and
HN$^{13}$C ($J=3\rightarrow2$) were modelled in a similar fashion
to obtain their excitation temperatures, since these were the only
credible transitions detected for these molecules.

The $^{13}$C$^{16}$O lines are optically thick, as is made clear
from their self-absorbed profiles (see Figure \ref{fig3}). In order
to obtain the excitation temperature for $^{13}$C$^{16}$O, we took
the ratio of the integrated intensities in the wings of the line profiles,
where the line optical depth is small ($\beta\sim1$) and applied
the method described above. 

Finally, the column densities of all molecules, except $^{13}$C$^{16}$O
and $^{12}$C$^{16}$O, were calculated using equation (\ref{eq:Nt})
and the values of $\tau$ and $T_{ex}$ as explained above. These
results are summarized in Table \ref{ta:tex} (the cases of $^{13}$C$^{16}$O
and $^{12}$C$^{16}$O are considered in the next section). The excitation
temperatures obtained for the different species fall in a narrow range
of 12 to 22 K, which is consistent with the assumption of LTE and
the kinetic temperature of 20 K discussed in $\S 3$.

\placetable{ta:tex}

\subsection{C and O isotopic ratios}

The $^{12}$C/$^{13}$C isotopic ratio was directly determined from
the ratio of the column density of $^{12}$C$^{18}$O to that of $^{13}$C$^{18}$O.
The observation of these relatively low-abundance species has the
advantage of probing denser regions over the observation of more abundant
isotopologues such as $^{12}$C$^{16}$O and $^{13}$C$^{16}$O. 

For the carbon isotopic ratio, we obtain 

\begin{equation}
\frac{^{12}\mathrm{C}}{^{13}\mathrm{C}}\equiv\frac{N_{tot}(\mathrm{^{12}C^{18}O})}{N_{tot}(\mathrm{^{13}C^{18}O})}=63\pm4,\label{eq:c12/c13}\end{equation}

\noindent which is consistent with the results of \citet{lp93} in
other clouds. 

In order to determine the $^{16}$O/$^{18}$O isotopic ratio, it was
necessary to compare the abundance of $^{13}$C$^{16}$O with that
of $^{13}$C$^{18}$O. Since the $^{13}$C$^{16}$O profiles show
self-absorption, we took the intensity ratio, $R$, of the $J=2\rightarrow1$
transitions of $^{13}$C$^{16}$O and $^{12}$C$^{18}$O in a common
velocity interval in the wings, where the line opacities are relatively
small (we included the optical depth correction, $\beta$, for $^{13}$C$^{16}$O
in this ratio). There is an uncertainty in the obtained value for
the column density of $^{13}$C$^{16}$O, since $R$ may vary over
the line profile. Nevertheless, we examined this ratio in two different
parts of each wing in order to check its consistency. The $^{16}$O/$^{18}$O
isotopic ratio is determined to be

\begin{equation}
\frac{^{16}\mathrm{O}}{^{18}\mathrm{O}}\equiv\beta R\Bigl(\frac{^{12}\mathrm{C}}{^{13}\mathrm{C}}\Bigr)=318\pm64,\label{eq:o16/o18}\end{equation}

\noindent which is in agreement with previous studies in other clouds
(\citealt{penzias}; \citealt{Polehampton}). 

The column densities of the species listed in Table \ref{ta:tex}
were then calculated from their isotopologues using the obtained C
and O isotopic ratios. We determined the column density of $^{13}$C$^{16}$O
and $^{12}$C$^{16}$O and estimated the column density of molecular
hydrogen in DR21(OH) to be \citep{stahler} 

\begin{equation}
N\mathrm{(H}_{2})\simeq3.8\times10^{5}N(^{13}\mathrm{CO)}\simeq1.1\times10^{23}\:\mathrm{cm^{-2}}.\label{eq:N_H2}\end{equation}

\subsection{Fractional ionization and cosmic ray ionization rate}

We determined the H$_{3}$$^{+}$ and electron abundances for DR21(OH)
using the column density equivalent of equations (\ref{eq:n(e)1})
and (\ref{eq:h3+}), and equation (\ref{eq:N_H2}). For the fractional
ionization and H$_{3}$$^{+}$ column density we report $\chi_{\mathrm{e}}=3.2\times10^{-8}$
and $N(\mathrm{H}_{3}^{+})=5.5\times10^{13}\:\mathrm{cm^{-2}}$, respectively.
\citet{mccall} detected absorption lines of H$_{3}$$^{+}$ in several
dense molecular clouds, and estimated $N(\mathrm{H}_{3}^{+})$ to
be $(1-5)\times10^{14}$$\:\mathrm{cm^{-2}}$. We find our calculated
value for $N(\mathrm{H}_{3}^{+})$ lower than theirs. Our result for
$\chi_{\mathrm{e}}$ agrees well with previous findings in other sources
\citep{bergin,flower}. Moreover, we estimate the cosmic ray ionization
rate for DR21(OH)Main to be $\zeta_{\mathrm{H_{2}}}=3.1\times10^{-18}\:\mathrm{s}^{-1}$
through equation (\ref{eq:zeta}), which agrees with recent results
for other clouds (e.g., \citealt{flower}). 

It should be noted that the rate coefficients for the chemical reactions
are either calculated or measured in laboratories, and have uncertainties
ranging from $25$$\%$ up to a factor of 2 \citep{Woodall}. The
OSU database has historically been used for chemical modelling of
dark clouds; however, the latest release of the UMIST database contains
dipole reactions with fits specific for low temperatures and is well-suited
to this type of modelling. The difference in the rate coefficients
between the two databases are within a factor of two and, in order
to investigate the effect of this, we repeated our calculations using
the rate coefficients from the OSU database. We find an increase in
the ionization fraction and the cosmic rate ionization rate by a factor
of two, which is commensurate with the uncertainties in the rate coefficients
mentioned above. 

The cloud size along the line of sight is needed for the calculation
of the molecular number densities, and can be approximately estimated
through the inspection of extended emission maps of molecules in the
cloud. But the uncertainty in the distance of the source adds to the
inaccuracy in the determination of the cloud size. We estimated the
cloud size from the HCN ($J=4\rightarrow3$) map by \citet{richardson2}
to be 1.7 pc. Finally, the column density of molecular hydrogen in
dense clouds is usually calculated using standard ratios of CO/H$_{2}$
or through extinction measurements. All the uncertainties in the aforementioned
parameters cause $\zeta_{\mathrm{H_{2}}}$ to be uncertain by a factor
of a few.

\section{Discussion and Summary}

We have presented a simple yet novel method to simultaneously estimate
the ionization fraction and cosmic ray ionization rate of H$_2$ in
dense molecular clouds. These parameters were determined in DR21(OH)
using ancillary observational data pertaining to the formation and
destruction reactions of HCNH$^+$ and HCO$^+$, which involve the
H$_3$$^+$ and electron abundance. We have made a number of assumptions
regarding this method, which will be discussed here.

a) The primary assumption in this method is that cosmic rays are the
dominant means of ionization in DR21(OH). As mentioned earlier, there
has been no detection of HII regions in DR21(OH), which implies there
are no actively photo-ionizing stars in the cloud. \citet{gibb} mapped
the DR21/DR21(OH) region at 3 mm, 850 $\mu\mathrm{m}$ and 450 $\mu\mathrm{m}$
and identified several submillimeter sources within the region, which
only suggest the existence of deeply embedded young protostars in
the source. We note that the value we obtain for the ionization fraction
is consistent with our assumption.

b) We selected the most significant formation and destruction routes
for HCNH$^+$ and HCO$^+$ by identifying the reactions that had the
largest rate coefficients and involved significantly abundant molecules.
There are other ionic and neutral species such as CO$^+$, C$^+$,
HCN$^+$, and H$_2$O that take part in the HCNH$^+$ and HCO$^+$
chemistry. These species are abundant in photon dominated regions
\citep{savage}, but can safely be neglected here given the lack of
photo-ionizing sources in DR21(OH), as noted above. 

c) For our calculations, we have assumed chemical equilibrium within
the region under study. \citet{lintott} argue that in a rapidly evolving
cloud, the dynamical timescale may be shorter than the chemical timescale,
and hence chemical equilibrium will not be established until the cloud
has reached a quiescent stage, i.e., passed its initial collapse phase.
They mention this timescale to be on the order of $10^{6}$ years.
If chemical equilibrium is not met in DR21(OH), then $\zeta_{\mathrm{H_{2}}}$
will be overestimated by a factor of $2-3$ \citep{lintott}. This
uncertainty is comparable with the uncertainties in the rate coefficients,
observational calibrations, and measurements of cloud size and molecular
hydrogen abundance. 

d) DR21(OH) is known to harbor water and methanol masers (associated with the MM1 and MM2 continuum sources), being  an indication of the presence of outflows within the source. \citet{magnum, magnum2} observed this source using the VLA and found an excitation temperature above 80 K within a region of ² $\sim10\arcsec$ centered on MM1 (VLSR $\sim$ -4.1 km s$^{-1}$). Gas associated with such regions would understandably have different physical characteristics compared to the more extended, relatively quiescent regions probed with our observations. It is the latter that dominates the region we observed with our larger telescope beam ($\sim20\arcsec$ to $\sim30\arcsec$). Nevertheless, previous studies have established that this star-forming region possesses a complex structure owing to the presence of deeply embedded sources, and the detection of molecular masers further supports this conclusion. One could thus question our assignment of a single kinetic temperature of 20 K for DR21(OH). But since we obtained a narrow range of temperatures (12 - 22 K) for all the molecular species we observed, we believe that our aforementioned chosen value is justified for the spatial resolution attained with our observations. Incidentally, this narrow range of temperatures for all species is also consistent with the use of the LTE approximation for this study.

Provided the above assumptions are met, our method will be very useful
for the calculation of $\chi_{\mathrm{e}}$ and $\zeta_{\mathrm{H_{2}}}$
in dense molecular clouds without the need for detailed chemical models.
However, the application of this technique can be limited due to the
likely difficulty in detecting HCNH$^+$. For example, we attempted
to observe HCNH$^+$ in other molecular clouds such as W3(OH), AFGL2591
and AFGL490 but could not record any credible detection. For these
situations, other molecules such as HCS$^+$, whose formation and
destruction follow a similar chemistry and could be coexistent with
HCO$^+$ may be taken into consideration. 

Moreover, it is not straightforward to precisely determine the density for which our values of the ionization fraction and cosmic ray ionization rate apply when one considers that our array of observations involves molecules with a vast range of critical densities (e.g., from $\sim$ 10$^{3}$ cm$^{-3}$ for CO to $\sim$ 10$^{8}$ cm$^{-3}$ for HCN). We had to assume that all observed species are coexistent within DR21(OH) in order to carry our program; however, this is probably not strictly accurate.  This is especially the case for CO and its isotopologues, because of their lower critical densities, and this brings a further uncertainty in our determination of $\chi_{\mathrm{e}}$ and $\zeta_{\mathrm{H_{2}}}$. Unfortunately, it is difficult to evaluate the magnitude of this bias with our existing data. On the other hand, the good alignment of spectral peak intensities and the relative agreement on excitation temperatures is consistent with our coexistence approximation.     

Although our method does not take account of gas-grain interactions
or deuterated species, which may be important for clouds at very low
temperatures (e.g. \citealt{flower}), its simplicity and reliance
on observations render it a powerful tool for the simultaneous calculation
of $\chi_{\mathrm{e}}$ and $\zeta_{\mathrm{H_{2}}}$. Moreover, this
method enables us to indirectly estimate the column density of H$_{3}$$^{+}$
in dense molecular clouds. This is important, since the low abundance
of H$_{3}$$^{+}$ makes its absorption lines difficult to observe,
and it is not at all detectable in the submillimeter regime. Furthermore,
the obtained values for $\chi_{\mathrm{e}}$ and $\zeta_{\mathrm{H_{2}}}$
are applicable in the calculations of ambipolar diffusion timescale
and the magnitude of the mean magnetic field in the cloud, which will
be addressed in a future paper.

\acknowledgements{The authors thank the anonymous referee for a careful reading and insightful comments, and also D. Johnstone and J. Di Francesco for helpful
discussions and suggestions. M. H.'s research is funded through the NSERC
Discovery Grant, Canada Research Chair, Canada Foundation for Innovation,
Ontario Innovation Trust, and Western's Academic Development Fund
programs. The Caltech Submillimeter Observatory is funded through
the NSF grant AST 05-40882 to the California Institute of Technology. }

\clearpage

\begin{deluxetable}{lclrrrrr}

\tabletypesize{\footnotesize}

\tablecaption{Data for the observed molecular transitions  \label{ta:data}}

\tablecolumns{8}

\tablewidth{0pt}

\tablehead{




\colhead{Molecular transition} & \colhead{Frequency} & 

\colhead{A$_{ul}$} & \colhead{T$_{peak}$} & \colhead{V$_{peak}$} & \colhead{$\int TdV$}  & \colhead{Velocity range} &\colhead{$\tau$} \\

\colhead{} & \colhead{GHz} & \colhead{s$^{-1}$}&\colhead{K} & \colhead{km s$^{-1}$} & \colhead{K km s$^{-1}$}  & \colhead{km s$^{-1}$} &\colhead{} \\  

}

\startdata

HCNH$^{+}$ (3-2) & 222.329 & 4.61$\times$10$^{-6}$ & 0.062 & -3.13 & 0.32 $\pm$ 0.02   &(-6.1, 0.3) & 0.003 \\

HCNH$^{+}$ (4-3) & 296.433 & 1.13$\times$10$^{-5}$ & 0.064 & -3.27 & 0.29 $\pm$ 0.02   & (-6.1, -1.3) & 0.004 \\

H$_{3}$O$^{+}$ (3$_{2}$-2$_{2}$) & 364.797 & 2.79$\times$10$^{-4}$ & 0.16 & -3.16 & 1.38 $\pm$ 0.08 &(-6.2, 1.2) & 0.02 \\

HN$^{13}$C (3-2) & 261.263 & 6.48$\times$10$^{-4}$ & 0.38 & -3.07 &  2.99 $\pm$ 0.03   &(-6.7, 1.1)& 0.06\\

HN$^{13}$C (4-3)\tablenotemark{a} & 348.340  & 1.59$\times$10$^{-3}$ & 0.66 & -3.81 & 6.14 $\pm$ 0.08   &(-10.0, 1.0)& \nodata\\

H$^{13}$CN (3-2) & 259.011 &  7.72$\times$10$^{-4}$  & 0.96 & -2.84 & 8.67 $\pm$ 0.03    &(-9.4, 3.8)& 0.09\\

H$^{13}$CN (4-3) & 345.339 & 1.9$\times$10$^{-3}$ & 0.39 & -2.70 & 4.17 $\pm$ 0.02   &(-7.6, 3.5)& 0.06\\

H$^{13}$CO$^{+}$ (3-2) & 260.255 &  1.34$\times$10$^{-3}$  & 1.02 & -2.94 & 8.37 $\pm$ 0.03    &(-7.4, 2.1)& 0.09\\

H$^{13}$CO$^{+}$ (4-3) & 346.998 &  3.29$\times$10$^{-3}$ & 0.56 & -2.70 & 4.89 $\pm$ 0.02     &(-7.2, 1.8)& 0.07\\

$^{13}$C$^{18}$O (2-1) & 209.419 & 5.23$\times$10$^{-6}$ & 0.20 & -3.00 & 0.91 $\pm$ 0.06    &(-6.1, 0.0)& 0.01\\

$^{13}$C$^{18}$O (3-2) & 314.119 & 1.89$\times$10$^{-6}$ & 0.22 & -3.14 & 0.90 $\pm$ 0.03  &(-5.9, -0.9)& 0.01\\

$^{12}$C$^{18}$O (2-1) & 219.560 & 6.01$\times$10$^{-7}$ & 6.60 & -3.14 & 41.44 $\pm$ 0.07   &(-7.0, 1.7)& 2.19\tablenotemark{c}\\

$^{12}$C$^{18}$O (3-2)\tablenotemark{a} & 329.330 & 2.17$\times$10$^{-6}$ & 6.06 & -3.70 & 40.28 $\pm$ 0.12    &(-7.0, 0.6)& \nodata\\
$^{13}$C$^{16}$O (2-1)\tablenotemark{b} & 220.398 & 6.07$\times$10$^{-7}$ & \nodata & \nodata & 92.70 $\pm$ 0.14    &(-8.2, 2.8) & \nodata\\
$^{13}$C$^{16}$O (3-2)\tablenotemark{b} & 330.587 & 2.19$\times$10$^{-6}$ & \nodata & \nodata & 40.45 $\pm$ 0.32    &(-7.8, 1.9) & \nodata\\

\enddata


\tablenotetext{a}{Not used in the analysis due to shifted V$_{peak}$.}

\tablenotetext{b}{Two velocity components, shows strong self-absorption.}

\tablenotetext{c}{optical depth in the center of the line.}
\end{deluxetable}

\clearpage

\begin{deluxetable}{cll}

\tabletypesize{\footnotesize}

\tablecaption{Rate coefficients\tablenotemark{a} of the reactions for H$_{3}$$^{+}$, HCNH$^{+}$ and HCO$^{+}$  \label{ta:ki}}

\tablecolumns{3}

\tablewidth{0pt}

\tablehead{




\colhead{Index\tablenotemark{b} (i)} & \colhead{reaction} & \colhead{k$_{i}$ (cm$^{3}$ s$^{-1}$)} \\

}

\startdata

1 & H$_{3}$$^{+}$ + HCN(HNC) $\longrightarrow$ HCNH$^{+}$ + H$_{2}$ & 8.1$\times$10$^{-9}$(T/300 K)$^{-0.5}$    \\

2 & HCO$^{+}$ + HCN(HNC) $\longrightarrow$ HCNH$^{+}$ + CO & 3.1$\times$10$^{-9}$(T/300 K)$^{-0.5}$        \\

3 & H$_{3}$O$^{+}$ + HCN(HNC) $\longrightarrow$ HCNH$^{+}$ + H$_{2}$O  & 4.0$\times$10$^{-9}$(T/300 K)$^{-0.5}$       \\

4 & H$_{3}$$^{+}$ + CO $\longrightarrow$ HCO$^{+}$ + H$_{2}$& 1.7$\times$10$^{-9}$      \\

5\tablenotemark{c} & HCNH$^{+}$ + e$^{-}$ $\longrightarrow$ (HCN + H) or (HNC + H) or (CN + H + H) & 2.8$\times$10$^{-7}$(T/300 K)$^{-0.65}$        \\
  
6 & HCO$^{+}$ + e$^{-}$ $\longrightarrow$ CO + H & 2.4$\times$10$^{-7}$(T/300 K)$^{-0.69}$       \\

10 & H$_{2}$$^{+}$ + H$_{2}$ $\longrightarrow$ H$_{3}$$^{+}$ + H & 2.1$\times$10$^{-9}$   \\
11\tablenotemark{c} & H$_{3}$$^{+}$ + e$^{-}$ $\longrightarrow$ (H$_{2}$ + H) or (H + H + H) & 6.7$\times$10$^{-8}$(T/300 K)$^{-0.52}$\\
  
\enddata

\tablenotetext{a}{All the rate coefficients are taken from the UMIST database.}
\tablenotetext{b}{The index follows the equation number where the corresponding reaction first appears in the text.}
\tablenotetext{c}{The overall rate coefficient of this reaction is the sum of the rates of all the given branch reactions.}

\end{deluxetable}

\clearpage

\begin{deluxetable}{lcr}

\tabletypesize{\footnotesize}

\tablecaption{Excitation temperatures and column densities  \label{ta:tex}}

\tablecolumns{6}

\tablewidth{0pt}

\tablehead{




\colhead{Molecular Species} & \colhead{T$_{ex}$(K)} & \colhead{N$_{tot}$(cm$^{-2}$)} \\

}

\startdata

HCNH$^{+}$  & 22 & 3.2$\times$10$^{13}$    \\

H$_{3}$O$^{+}$  & 16 & 1.3$\times$10$^{16}$    \\

HN$^{13}$C  & 12 & 4.4$\times$10$^{12}$   \\

H$^{13}$CN  & 13 & 9.8$\times$10$^{12}$     \\

H$^{13}$CO$^{+}$  & 15 & 4.8$\times$10$^{12}$    \\

$^{13}$C$^{18}$O  & 20 & 9.0$\times$10$^{14}$   \\

$^{12}$C$^{18}$O  & 16 & 5.7$\times$10$^{16}$  \\

$^{13}$C$^{16}$O  & 18 & 2.9$\times$10$^{17}$   \\
$^{12}$C$^{16}$O  & \nodata & 1.8$\times$10$^{19}$   \\

\enddata



\end{deluxetable}

\clearpage

\end{document}